\documentclass[prl,twocolumn,showpacs]{revtex4}
\usepackage{graphicx}
\usepackage{amssymb}
\usepackage{epstopdf}
\usepackage[usenames]{color}
\begin{document}

\title{Coalescence in low-viscosity liquids}
\author{Sarah C. Case and Sidney R. Nagel}
\affiliation{The James Franck Institute and Department of Physics, University of Chicago, Chicago, Illinois 60637}

\begin{abstract}
The expected universal dynamics associated with the initial stage of droplet coalescence are difficult to study visually due to the rapid motion of the liquid and the awkward viewing geometry.  Here we employ an electrical method to study the coalescence of two inviscid droplets at early times.  We measure the growth dynamics of the bridge connecting the two droplets and observe a new asymptotic regime inconsistent with previous theoretical predictions.  The measurements are consistent with a model in which the two liquids coalesce with a slightly deformed interface.


\end {abstract}

\maketitle

When fluid drops merge, a dramatic transformation occurs: the topology changes as the fluid masses, originally separated, merge into a single entity.  At first, the drops are separated by only a small distance.  Then a thin fluid bridge is formed between them which rapidly widens due to surface tension forces.  We employ an electrical method to explore drop coalescence in a low viscosity fluid and find that, shortly after its initiation, there is an unexpected new regime that is dominated by the overall deformability of the drops.

Soon after the instant of coalescence, the length scales that characterize the fluid bridge are many orders of magnitude smaller than the macroscopic dimensions of the flow.  Such a separation of length scales often leads to universal behavior as in the reverse process to coalescence, drop break up, in which one drop breaks up into two or more droplets \cite{Shi_1994, Eggers_1997, Cohen_2001, Lister_1998, Chen_2002}.  In this case, the transition proceeds as the radius of the liquid neck connecting the two pieces of fluid approaches zero.  This causes the dynamics to approach a singularity, motivating the comparison of such fluid transitions to critical thermodynamic phase transitions\cite{Constantin_1993, Goldstein_1993, Bertozzi_1996}.  We also expect singular behavior during drop coalescence.  However, recent discoveries have shown that not all fluid singularities obey universal dynamics\cite{Doshi_2003, Keim_2006} so that fluid transitions have a broader range of behaviors than the analogy with critical phenomena might suggest.  

For low-viscosity fluids such as water, the widening of the bridge during coalescence is opposed primarily by inertial rather than viscous forces.  Thus, throughout much of coalescence, the bridge radius $r$ is much larger than the viscous length scale of water, $ l_{\nu} = \mu^2/\rho \gamma \approx 14$ nm.   A straightforward scaling argument has been used to describe the process \cite{Eggers_1999}.   
 
\begin{figure}
\centering
\includegraphics[width=90 mm]{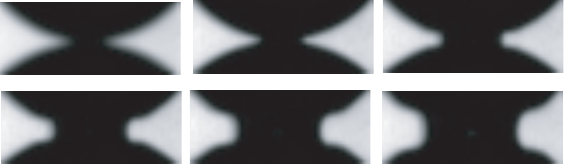}
\caption{Sequence of images showing the formation of a fluid bridge between two drops of aqueous NaCl solution at saturation,  fluid density $\rho = 1.1972$ g/cm, kinematic viscosity $\nu = 1.662$ cSt, and surface tension $\gamma = 82.55$ dyne/cm\cite{Handbook}.  The drop radius $A = 1$ mm.  The frames are separated by $69$ $\mu$s.} 
\end{figure} 

The film of air between the drops ruptures to form a fluid bridge only when the drops are separated by a small distance.  Soon after the bridge is formed, the length of the bridge, $d$, is far smaller than its radius, $r$.  Thus, $d$ is the relevant length scale for calculating the pressure due to interfacial tension.  Comparing interfacial tension to inertia, we find
\begin{equation}
 (\frac{\gamma}{\rho d})^{1/2} = \frac{dr}{dt}.
\end{equation}
For a bridge spreading between two static, hemispherical drops, $d = 2 r^{2}/A$, where $A$ is the drop radius.  The solution of this differential equation gives:
\begin{equation}
 r =  (\frac{4 \gamma A}{\rho})^{1/4} (t - t_{0})^{1/2} =  (\frac{4 \gamma A}{\rho})^{1/4} \tau^{1/2},
\end{equation}
where $t_0$ is the instant of coalescence and $\tau \equiv t - t_0$.

Simulations studying the coalescence of low-viscosity fluid drops in vacuum have confirmed this scaling law\cite{Duchemin_2003, Lee_2006}.  Also, experiments using high speed imaging at up to $10^6$ frames per second have observed $r \propto \tau^{1/2}$ for  $\tau > 10$ $\mu$s\cite{Thoroddsen_2005, Menchaca_2001, Wu_2004}.  However, the speed and the geometry of the transition limit the range of imaging studies so that times earlier than 10 $\mu$s were impossible to access.  Here we present an electrical method that allows us to study the fluid bridge between two coalescing drops at much shorter times: $\tau \sim 10$ ns.  An electrical method was developed by Burton et al., where a small DC voltage was used to measure the resistance of a drop of mercury during break up\cite{Burton_2004}.  Our method adapts that technique to an AC voltage.  This allows us to measure both the time-dependent resistance and capacitance of two coalescing drops of a conducting ionic solution such as aqueous NaCl.  We find a new regime when $\tau <  10 \mu$s inconsistent with the predictions of the scaling argument outlined above. 

\begin{figure}
\centering
\includegraphics[width=90 mm]{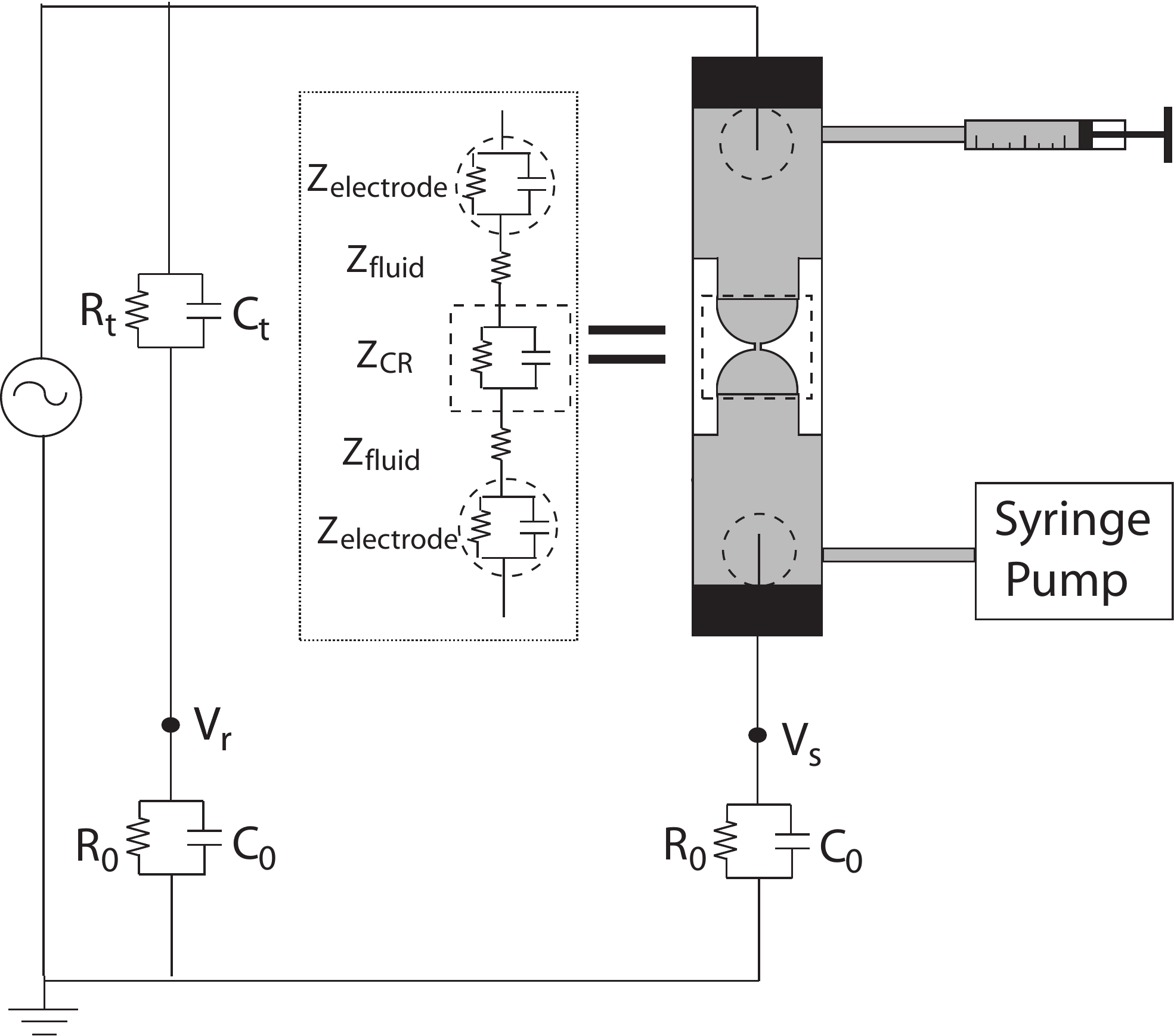}
\caption{Experimental Setup. Two nozzles of radius $A$ are secured in line with one another with tips separated by $2 A$.  A drop of aqueous NaCl solution is formed on the upper nozzle using a microliter syringe.  The lower drop is slowly grown until the two drops coalesce. Electrodes are secured in the measurement cell and connected to a Wheatstone bridge circuit.  The upper left branch consists of known circuit elements ($R_t$ and $C_t$), while each lower branch is the input impedance of an NI PCI-5105 high-speed simultaneous sampling digitizer ($R_0$) in parallel with the capacitance of the coaxial cables ($C_0$).  The impedance of the cell can be separated into three contributions added in series:  $Z_{electrodes}$, $Z_{fluid}$, and $Z_{CR}$.  $Z_{electrode}$ can be modeled as a frequency-dependent capacitance in parallel with an equivalent resistance due to charge transfer, as shown in the dotted circle\cite{Bockris}.  The impedance of the coalescing region, $Z_{CR}$, can be modeled as a capacitor in parallel with a resistor, as shown in the dashed square.  The conductivity of salt water at saturation is $\sigma = 0.225$ $(\Omega \cdot$ cm$)^{-1}$\cite{Handbook}
} 
\end{figure} 

In our experiments, two nozzles of inner radius $A$ were aligned vertically as shown in Fig. 2.  We filled the nozzles with saturated aqueous NaCl solution and formed a hemispherical drop at the tip of the upper nozzle by injecting a known quantity of fluid.  We applied a fixed amplitude AC voltage of frequency $f$ across electrodes secured opposite the nozzle tips, then grew the lower drop at a fixed rate until coalescence occurred.  As shown in Fig. 2, we used a Wheatstone bridge to measure $Z_{cell}$ as a function of time during coalescence.  We sampled two voltages simultaneously, $V_r$ and $V_s$.  With the exception of $f = 10$ MHz, which was sampled at the maximum rate of $60$ MHz, the sampling was done at $10f$.  Using Labview (National Instruments), we measured the ratio of the amplitudes, $|V_{r}|/|V_{s}|$ and the phase shift between them, $\Delta \phi$ as a function of time.  This allows us to calculate the real and imaginary parts of $Z_{cell}$.  We determined $t_0 \pm 1/f$ using the shift in $\Delta \phi$ at contact. 

This cell impedance can be separated into three contributions added in series, as seen in Fig. 2: 
\begin{equation}
Z_{cell} = 2 Z_{electrode} + 2 Z_{fluid} + Z_{CR}.
\end{equation} 
We model the electrode impedance $Z_{electrode}$ using equivalent circuit elements.  For a voltage across the cell $|V_{cell}| \lesssim 50 mV$, the charge transfer from the electrodes into the solution can be modeled as an equivalent resistance.  Additionally, the polarization of the fluid around the electrodes can be modeled as a frequency-dependent capacitor in parallel with this resistance\cite{Bockris}.  The fluid between the electrodes and the nozzle tips contributes an impedance $Z_{fluid}$, and the ``coalescing region'' between the tips of the nozzles contributes an impedance $Z_{CR}$.

We bring the nozzle tips into contact to measure $ Z_{closed} = 2Z_{electrodes} + 2Z_{fluid}$.  Thus, we isolate $Z_{CR} = Z_{cell} - Z_{closed}$.  We model $Z_{CR}$ as a resistor $R_{CR}$ (representing the resistance of the liquid in the shape formed by the two drops and the neck between them) in parallel with a capacitor $C_{CR}$ (representing the capacitance between the drop surfaces). These quantities depend on the geometry of the coalescing region and are time-dependent. Our experiment measures $R_{CR}$ and $C_{CR}$.

\begin{figure}
\centering
\includegraphics[width=80 mm]{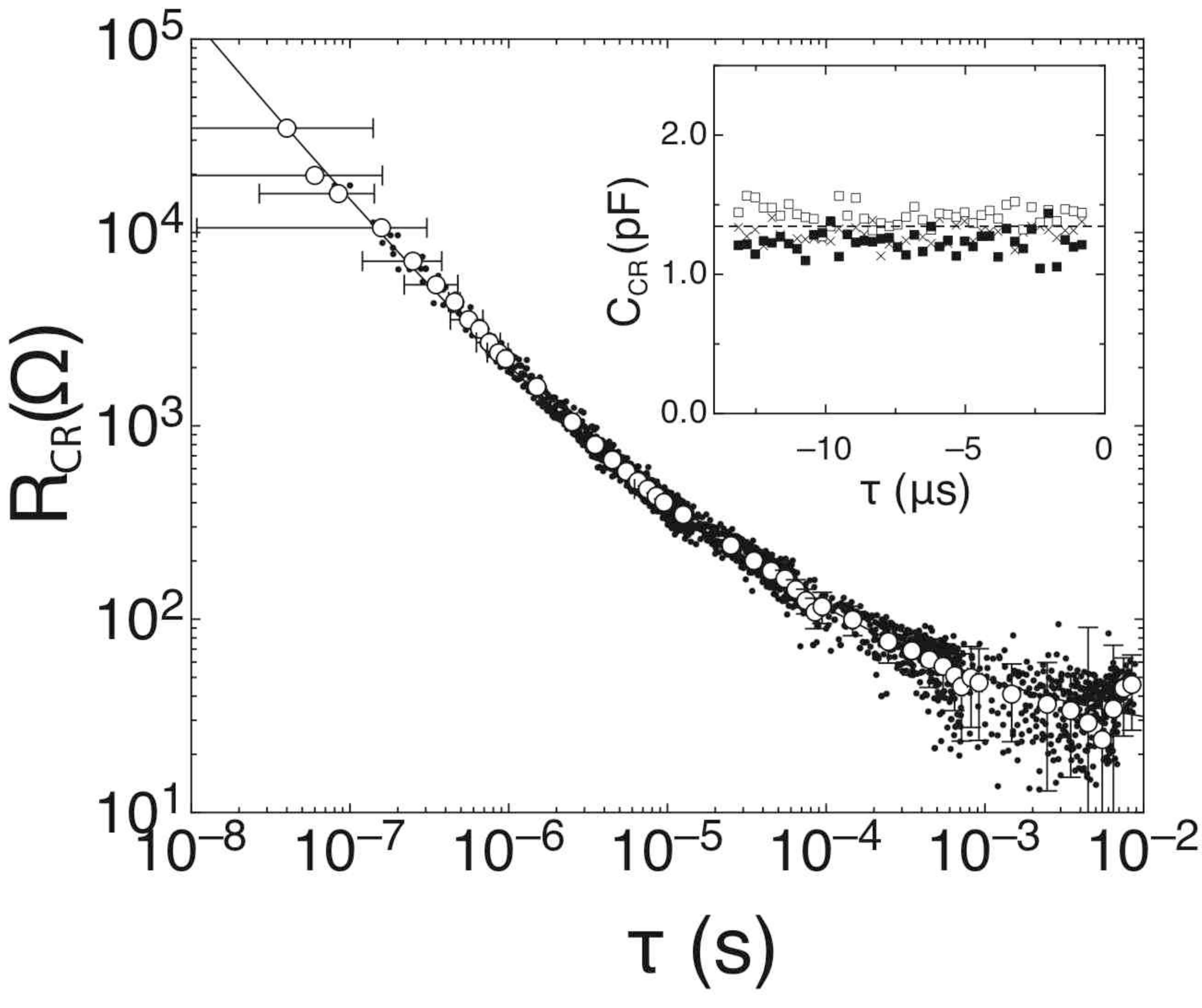}
\caption{Resistance during droplet coalescence.  $R_{CR}$ versus $\tau = (t - t_0)$.   $A  = 1$ mm. The drops approach one another at a rate of 0.0004 $A$/ms.  The closed symbols show 24 individual coalescence events, six obtained at each of four measurement frequencies.  The open symbols are the average of the closed symbols, binned logarithmically.  The error bars reflect the spread in these measurements as well as systematic error due to inaccuracies in the measurement of $Z_{electrode}$ and due to the choice of $t_0$, the instant of coalescence.  The solid line shows  $R_{CR} = 1.23\cdot 10^{-3})\tau^{-1} + 0.79 \tau^{-1/2} + 23.3$. The inset shows $C_{init}$ versus $\tau$.  Data is shown for three independent coalescence events at $f = 10$ MHz.  The dashed line is the average value of the data shown, $C_{init} = 1.3$ pF.} 
\end{figure}
	
$R_{CR}$ versus $\tau$ is shown in Fig. 3 for $A = 1$ mm.  The filled symbols show twelve independent coalescence events, where three events are taken at each of four frequencies $f$ ranging from $10$ kHz to $10$ MHz.  The open symbols show this data binned logarithmically and averaged.  We find that the data is described well by: $R_{CR} = \alpha \tau^{-1} + \beta \tau^{-1/2} + \delta$.  The solid line in Fig. 3 shows this fit with $\alpha=(1.23 \pm 0.3) \cdot 10^{-3}$, $\beta=0.8 \pm 0.2$ and $\delta=23 \pm 15$. 

The inset to Fig. 3 shows $C_{CR}$, the capacitance of the system during the $5$ $\mu$s before coalescence occurs for three independent coalescence events taken at $f = 10$ MHz.  For $f < 10$ MHz, the signal before coalescence is comparable to the system noise.  $C_{init}$ is constant within error, and the dashed line shows the average value, $C_{CR} = 1.3 \pm 0.14$ pF.  After coalescence, once the bridge is formed, the capacitance is poorly defined.   

We predict $R_{CR}$ and $C_{init}$, the capacitance of the coalescing region just before coalescence occurs, using the scaling argument of Eq. 2.  We separate $R_{CR}$ into three resistances connected in series,  $R_{CR} = R_{upper} + R_{bridge} + R_{lower}$.  $R_{upper}$ ($R_{lower}$) is the resistance between the upper (lower) nozzle tip and bridge, $R_{bridge}$ is the bridge resistance.

We model the drops such that $R_{upper} = R_{lower} = R_{hemi}$ where  $R_{hemi}$ is the resistance of a hemisphere truncated by a plane parallel to the flat surface of the hemisphere.  The plane intersects the hemisphere with radius $r_{tr}$, as shown in Fig. 4a.  The resistance of this shape can be calculated numerically using the electrostatics calculation package EStat (FieldCo).  We vary $r_{tr}$ over several orders of magnitude, and find $R_{hemi} =  1/4 \sigma r_{tr}$, where $\sigma$ is the conductivity of the fluid.    

For two hemispherical drops just touching at the tips, $d = 2 r^2/A$.  Therefore, as $R_{bridge}\approx  d/\sigma \pi r^2$,  we find that $R_{bridge} =  2/(\sigma \pi A)$, a constant.  

The scaling argument summarized by Eqn. (2) shows that $r \propto \tau^{1/2}$, and calculating $R_{CR}$ as a function of $\tau$, we find
\begin{equation}
R_{CR} =  \frac{1}{2 \sigma} (\frac{\rho}{4 \gamma A})^{1/4} \tau^{-1/2} + \frac{2}{\sigma \pi A}.
\end {equation}	

For aqueous NaCl solution in air and drop radius $A = 1$ mm,  $R_{CR} = 0.97 \tau^{-1/2} + 28.3$.  This result is in qualitative agreement with the data for $\tau > 10$ $\mu$s.  However, for $\tau \ll 10$ $\mu$s, $R_{CR} \sim \tau^{-1}$.  This is incompatible with the scaling argument.

\begin{figure}
\centering
\includegraphics[width=80 mm]{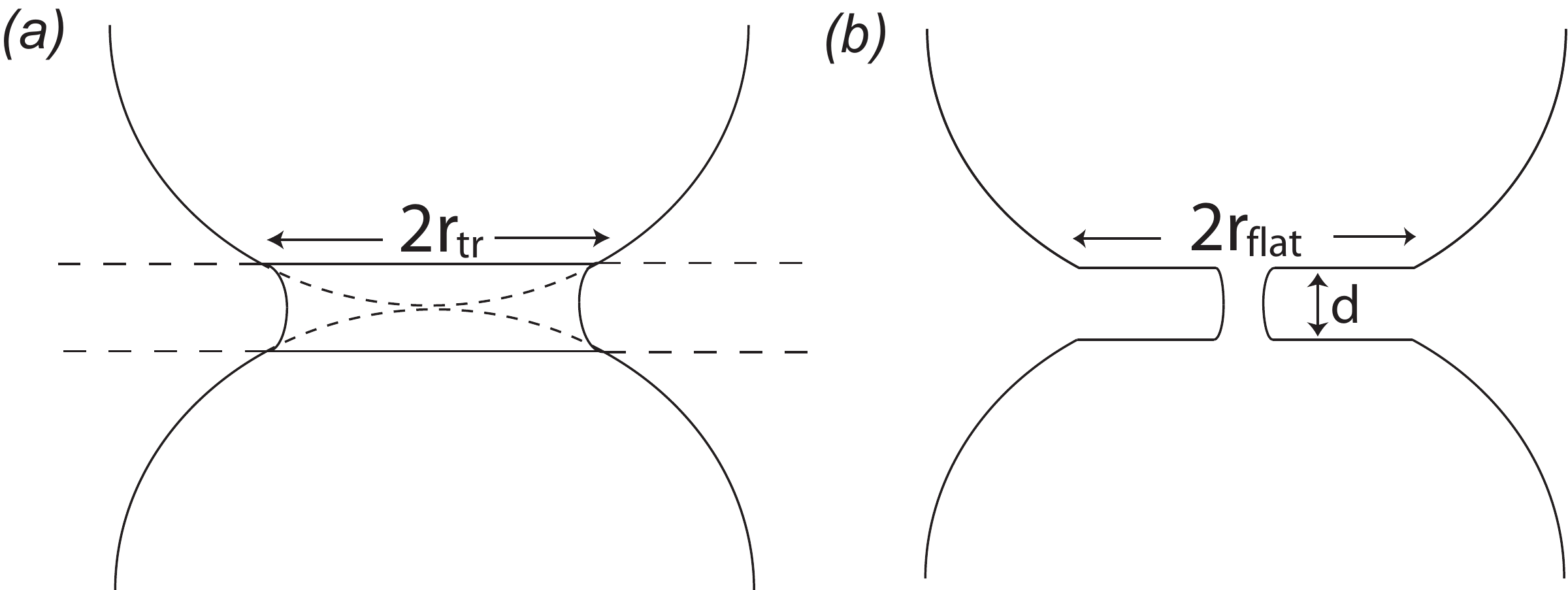}
\caption{ Two geometries for coalescence.  (a) Two hemispherical drops of radius $A$ coalesce, forming a bridge of maximum radius $r_{tr}$ and height $d = 2r_{tr}^2/A$.   (b)  Two drops coalesce with a flattened tip of  radius  $r_{flat}$.  Here, for a bridge of radius $r$, while $r < r_{flat}$, $d =$ constant.}
\end{figure}

We can account for this discrepancy with a slight modification of the geometry.  Eq. 2 was derived assuming $d \propto r^2$.  However, if the drop tips are not quadratic, this is no longer true.  For example, if prior to coalescence the drop tips are slightly flattened out to a radius $r_{flat}$, as shown in Fig. 4b, then at early times, $d$ is constant and only at late times would we see the $d \propto r^2$.  Assuming we remain in the inviscid regime, Eqn. (1) becomes 
\begin{equation}
        r = (\frac{\gamma}{\rho d})^{1/2} \tau.
\end{equation}

In order to calculate $R_{CR}$ in the altered geometry of Fig. 4b, rather than $R_{hemi}$, we model the resistance of a hemisphere with a hole of radius $r$ at the center of the flattened region radius $r_{flat}$.  We vary $r$, the radius of the rapidly widening bridge, over several orders of magnitude.  Solving numerically, we find that the resistance of this shape is $R_{flat} = 1/(4 r \sigma)$.  $R_{bridge}$ can be estimated as above, and for $d$ constant, we find that $R_{bridge} \approx  d/\sigma \pi r^2$.  We are able to find $R_{CR}$ versus time by combining this with Eq. 5:
\begin{equation}
R_{CR} = \frac{1}{2 \sigma} (\frac{\rho}{\gamma})^{1/2} \frac{d^{1/2}}{\tau} + \frac{\rho}{\sigma \pi \gamma} \frac{d^2}{ \tau^{2}}.
\end{equation}
This enables us to estimate the separation of the flattened tips, $d$, by comparing the predicted prefactor for the $\tau^{-1}$ term to our measurements.  We find $d = 200 \pm 100$ nm. 

In our experiments, there is a negligible contribution to $R_{CR}$ from the $\tau^2$ term.  Examining Eqn. (6), we see a crossover from $\tau^{-1}$ to $\tau^{-2}$ behavior at $\tau = 7 \pm 5$ ns for $d = 200 \pm 100$ nm.  Thus, we should see no contribution from the $\tau^{-2}$ term at the earliest times accessible by our experiments.

These conclusions are based on measurements of $R_{CR}$ versus $\tau$.  The inset to Fig. 3 shows our measurements of the capacitance $C_{CR}$.  However, the capacitance of two nearly-touching spheres is only logarithmically dependent on their separation \cite{Boyer_1994}.  Even a small error in our measurement of $C_{init}$ leads to enormous uncertainty in the calculated separation.  Moreover, since $C_{CR}$ represents the capacitance of the entire cell, the contributions from outside the ``coalescing region'' must be subtracted to obtain the relevant capacitance $C_{init}$, which represents the capacitance of only the drop tips \cite{TBP}.  We measure $C_{init} = 0.41 \pm 0.14$ pF.  This value is consistent with the model of the distorted drop tips but cannot exclude the original assumption of hemispherical drops at the instant of coalescence \cite{TBP}.

In conclusion, we have observed an unexpected asymptotic regime in the coalescence of two drops.  This regime becomes visible at $\tau < 10$ $\mu$s, which is earlier than the shortest times accessible by previous imaging studies.  Our electrical method allows us to study times three orders of magnitude earlier than this.  Previous experiments, theory and simulations indicated that drops coalesce while maintaining shapes described by quadratic minima.  However, our data is inconsistent with that picture and suggests that the coalescence occurs at the interface between two slightly flattened hemispherical drops.

The scaling arguments and simulations discussed earlier do not account for such a flattening, and an understanding of this phenomenon might illuminate other physical questions, such as the origin of the thin film rupture that initiates coalescence\cite{Oron_1997}.  Also, industrial applications of fluid drops, such as coating procedures, inkjet printing, and mixing in microfluidics could be affected by the way in which drops coalesce at the smallest scales and earliest times. 

Very near the instant of coalescence, the small-scale flows are decoupled from the large-scale flows.  Because of this, we expect the analogy between topological transitions in fluids and critical phase transitions would be most accurate at these earliest stages.  However, we find that the scaling in the drop coalescence transition is dependent on the geometry and deformability of the drops.  This has no analog in critical phase transitions.  			
		
	We are grateful to X. Cheng, E. Corwin, N. Keim, J. Royer, J. L. Wyman, W. Zhang, and L. N. Zou for helpful discussions.  This research was supported by NSF MRSEC DMR-0213745 and NSF DMR-0652269.


\begin{thebibliography}{16}
\expandafter\ifx\csname natexlab\endcsname\relax\def\natexlab#1{#1}\fi
\expandafter\ifx\csname bibnamefont\endcsname\relax
  \def\bibnamefont#1{#1}\fi
\expandafter\ifx\csname bibfnamefont\endcsname\relax
  \def\bibfnamefont#1{#1}\fi
\expandafter\ifx\csname citenamefont\endcsname\relax
  \def\citenamefont#1{#1}\fi
\expandafter\ifx\csname url\endcsname\relax
  \def\url#1{\texttt{#1}}\fi
\expandafter\ifx\csname urlprefix\endcsname\relax\def\urlprefix{URL }\fi
\providecommand{\bibinfo}[2]{#2}
\providecommand{\eprint}[2][]{\url{#2}}

        \bibitem[{\citenamefont{Shi et~al.}(1994)}]{Shi_1994}
\bibinfo{author}{\bibfnamefont{X.~D.}~\bibnamefont{Shi}},
\bibinfo{author}{\bibfnamefont{M.~P.}~\bibnamefont{Brenner}}, \bibnamefont{and}
  \bibinfo{author}{\bibfnamefont{S.~R.} \bibnamefont{Nagel}}, 
 \bibinfo{journal}{Science} \textbf{\bibinfo{volume}{265}},
  \bibinfo{pages}{219-222} (\bibinfo{year}{1994}).

\bibitem[{\citenamefont{Eggers}(1997)}]{Eggers_1997}
\bibinfo{author}{\bibfnamefont{J.}~\bibnamefont{Eggers}},
  \bibinfo{journal}{Rev. Mod. Phys} \textbf{\bibinfo{volume}{69}},
  \bibinfo{pages}{865} (\bibinfo{year}{1997}).

\bibitem[{\citenamefont{Lister and Stone}(1998)}]{Lister_1998}
\bibinfo{author}{\bibfnamefont{J.~R.} \bibnamefont{Lister}} \bibnamefont{and}
  \bibinfo{author}{\bibfnamefont{H.~A.} \bibnamefont{Stone}},
  \bibinfo{journal}{Phys. Fluids} \textbf{\bibinfo{volume}{10}},
  \bibinfo{pages}{2758} (\bibinfo{year}{1998}).

\bibitem[{\citenamefont{Cohen and Nagel}(2001)}]{Cohen_2001}
\bibinfo{author}{\bibfnamefont{I.}~\bibnamefont{Cohen}} \bibnamefont{and}
  \bibinfo{author}{\bibfnamefont{S.~R.} \bibnamefont{Nagel}},
  \bibinfo{journal}{Phys. Fluids} \textbf{\bibinfo{volume}{13}},
  \bibinfo{pages}{3533} (\bibinfo{year}{2001}).
  
   \bibitem[{\citenamefont{Chen et~al.}(2002)\citenamefont{Chen, Notz, and Basaran}}]{Chen_2002}
\bibinfo{author}{\bibfnamefont{A.~U.}~\bibnamefont{Chen}},
\bibinfo{author}{\bibfnamefont{P.~K.}~\bibnamefont{Notz}}, \bibnamefont{and}
  \bibinfo{author}{\bibfnamefont{O.~A.} \bibnamefont{Basaran}}, 
 \bibinfo{journal}{Phys. Rev. Lett.} \textbf{\bibinfo{volume}{88}},
  \bibinfo{pages}{174501} (\bibinfo{year}{2002}).

\bibitem[{\citenamefont{Constantin et~al.}(1993)\citenamefont{Constantin,
  Dupont, Goldstein, Kadanoff, Shelley, and Zhou}}]{Constantin_1993}
\bibinfo{author}{\bibfnamefont{P.}~\bibnamefont{Constantin}},
  \bibinfo{author}{\bibfnamefont{T.~F.} \bibnamefont{Dupont}},
  \bibinfo{author}{\bibfnamefont{R.~E.} \bibnamefont{Goldstein}},
  \bibinfo{author}{\bibfnamefont{L.~P.} \bibnamefont{Kadanoff}},
  \bibinfo{author}{\bibfnamefont{M.~J.} \bibnamefont{Shelley}},
  \bibnamefont{and} \bibinfo{author}{\bibfnamefont{S.~M.} \bibnamefont{Zhou}},
  \bibinfo{journal}{Phys. Rev. E} \textbf{\bibinfo{volume}{47}},
  \bibinfo{pages}{4169} (\bibinfo{year}{1993}).

\bibitem[{\citenamefont{Goldstein et~al.}(1993)\citenamefont{Goldstein, Pesci,
  and Shelley}}]{Goldstein_1993}
\bibinfo{author}{\bibfnamefont{R.~E.} \bibnamefont{Goldstein}},
  \bibinfo{author}{\bibfnamefont{A.~I.} \bibnamefont{Pesci}}, \bibnamefont{and}
  \bibinfo{author}{\bibfnamefont{M.~J.} \bibnamefont{Shelley}},
  \bibinfo{journal}{Phys. Rev. Lett.} \textbf{\bibinfo{volume}{70}},
  \bibinfo{pages}{3043} (\bibinfo{year}{1993}).

\bibitem[{\citenamefont{Bertozzi et~al.}(1996)\citenamefont{Bertozzi, Brenner,
  Dupont, and Kadanoff}}]{Bertozzi_1996}
\bibinfo{author}{\bibfnamefont{A.~L.} \bibnamefont{Bertozzi}},
  \bibinfo{author}{\bibfnamefont{M.~P.} \bibnamefont{Brenner}},
  \bibinfo{author}{\bibfnamefont{T.~F.} \bibnamefont{Dupont}},
  \bibnamefont{and} \bibinfo{author}{\bibfnamefont{L.~P.}
  \bibnamefont{Kadanoff}}, in \emph{\bibinfo{booktitle}{Trends and Perspectives
  in Applied Mathematics}}, edited by
  \bibinfo{editor}{\bibfnamefont{L.}~\bibnamefont{Sirovich}}
  (\bibinfo{publisher}{Springer (New York)}, \bibinfo{year}{1996}).

\bibitem[{\citenamefont{Doshi et~al.}(2003)\citenamefont{Doshi, Cohen, Zhang,
  Siegel, Howell, Basaran, and Nagel}}]{Doshi_2003}
\bibinfo{author}{\bibfnamefont{P.}~\bibnamefont{Doshi}},
  \bibinfo{author}{\bibfnamefont{I.}~\bibnamefont{Cohen}},
  \bibinfo{author}{\bibfnamefont{W.~W.} \bibnamefont{Zhang}},
  \bibinfo{author}{\bibfnamefont{M.}~\bibnamefont{Siegel}},
  \bibinfo{author}{\bibfnamefont{P.}~\bibnamefont{Howell}},
  \bibinfo{author}{\bibfnamefont{O.}~\bibnamefont{Basaran}}, \bibnamefont{and}
  \bibinfo{author}{\bibfnamefont{S.~R.} \bibnamefont{Nagel}},
  \bibinfo{journal}{Science} \textbf{\bibinfo{volume}{302}},
  \bibinfo{pages}{1185} (\bibinfo{year}{2003}).

\bibitem[{\citenamefont{Keim et~al.}(2006)\citenamefont{Keim, Moller, Zhang, and
  Nagel}}]{Keim_2006}
\bibinfo{author}{\bibfnamefont{N.}~\bibnamefont{Keim}},
\bibinfo{author}{\bibfnamefont{P.}~\bibnamefont{Moller}},
  \bibinfo{author}{\bibfnamefont{W.~W.} \bibnamefont{Zhang}}, \bibnamefont{and}
  \bibinfo{author}{\bibfnamefont{S.~R.} \bibnamefont{Nagel}},
 \bibinfo{journal}{Phys. Rev. Lett.} \textbf{\bibinfo{volume}{97}},
  \bibinfo{pages}{144503} (\bibinfo{year}{2006}).

  
 \bibitem[{\citenamefont{Eggers et~al.}(1999)\citenamefont{Eggers, Lister, and Stone}}]{Eggers_1999}
\bibinfo{author}{\bibfnamefont{J.}~\bibnamefont{Eggers}},
\bibinfo{author}{\bibfnamefont{J.~R.}~\bibnamefont{Lister}}, \bibnamefont{and}
  \bibinfo{author}{\bibfnamefont{H.~A.} \bibnamefont{Stone}}, 
 \bibinfo{journal}{J. Fluid. Mech.} \textbf{\bibinfo{volume}{401}},
  \bibinfo{pages}{293} (\bibinfo{year}{1999}).
  
  \bibitem[{\citenamefont{Thoroddsen et~al.}(2005)\citenamefont{Takahara and Etoh}}]{Thoroddsen_2005}
\bibinfo{author}{\bibfnamefont{S.~T.}~\bibnamefont{Thoroddsen}},
\bibinfo{author}{\bibfnamefont{K.}~\bibnamefont{Takahara}}, \bibnamefont{and}
  \bibinfo{author}{\bibfnamefont{T.~G.} \bibnamefont{Etoh}}, 
 \bibinfo{journal}{J. Fluid. Mech.} \textbf{\bibinfo{volume}{527}},
  \bibinfo{pages}{85} (\bibinfo{year}{2005}).

 \bibitem[{\citenamefont{Menchaca-Rocha et~al.}(2001)\citenamefont{Menchaca-Rocha, Mart\'{\i}nez-D\'avalos, N\'u\~{n}ez, Popinet, Zaleski}}]{Menchaca_2001}
\bibinfo{author}{\bibfnamefont{A.}~\bibnamefont{Menchaca-Rocha}},
\bibinfo{author}{\bibfnamefont{A.}~\bibnamefont{Mart\'{\i}nez-D\'avalos}},
\bibinfo{author}{\bibfnamefont{R.}~\bibnamefont{N\'u\~{n}ez}}, 
\bibinfo{author}{\bibfnamefont{S.}~\bibnamefont{Popinet}},  \bibnamefont{and}
  \bibinfo{author}{\bibfnamefont{S.} \bibnamefont{Zaleski}}, 
 \bibinfo{journal}{Phys. Rev. E} \textbf{\bibinfo{volume}{63}},
  \bibinfo{pages}{046309} (\bibinfo{year}{2001}).
 
   
     \bibitem[{\citenamefont{Wu et~al.}(2004)\citenamefont{Wu, Cubaud, and Ho}}]{Wu_2004}
\bibinfo{author}{\bibfnamefont{M.}~\bibnamefont{Wu}},
\bibinfo{author}{\bibfnamefont{T.}~\bibnamefont{Cubaud}}, \bibnamefont{and}
  \bibinfo{author}{\bibfnamefont{C.-M.} \bibnamefont{Ho}}, 
 \bibinfo{journal}{Phys. Fluids} \textbf{\bibinfo{volume}{16}},
  \bibinfo{pages}{L51} (\bibinfo{year}{2004}).
  
  \bibitem[{\citenamefont{Burton et~al.}(2004)\citenamefont{Burton, Rutledge, and Taborek}}]{Burton_2004}
\bibinfo{author}{\bibfnamefont{J.~C.}~\bibnamefont{Burton}},
\bibinfo{author}{\bibfnamefont{J.~E.}~\bibnamefont{Rutledge}}, \bibnamefont{and}
  \bibinfo{author}{\bibfnamefont{P.} \bibnamefont{Taborek}}, 
 \bibinfo{journal}{Phys. Rev. Lett.} \textbf{\bibinfo{volume}{92}},
  \bibinfo{pages}{244505} (\bibinfo{year}{2004}).
  
   \bibitem[{\citenamefont{Duchemin et~al.}(2003)\citenamefont{Duchemin, Eggers, and Josserand}}]{Duchemin_2003}
\bibinfo{author}{\bibfnamefont{L.}~\bibnamefont{Duchemin}},
\bibinfo{author}{\bibfnamefont{J.}~\bibnamefont{Eggers}}, \bibnamefont{and}
  \bibinfo{author}{\bibfnamefont{C.} \bibnamefont{Josserand}}, 
 \bibinfo{journal}{J. Fluid Mech.} \textbf{\bibinfo{volume}{487}},
  \bibinfo{pages}{167} (\bibinfo{year}{2003}).
  
   \bibitem[{\citenamefont{Lee et~al.}(2003)\citenamefont{Lee and Fischer}}]{Lee_2006}
\bibinfo{author}{\bibfnamefont{T.}~\bibnamefont{Lee}}, \bibnamefont{and}
\bibinfo{author}{\bibfnamefont{P.~F.}~\bibnamefont{Fischer}}, 
 \bibinfo{journal}{Phys. Rev. E} \textbf{\bibinfo{volume}{74}},
  \bibinfo{pages}{046709} (\bibinfo{year}{2006}).

  \bibitem[{\citenamefont{Bockris et~al.}(2004)\citenamefont{Bockris, Reddy, and Gamboa-Aldeco}}]{Bockris}
\bibinfo{author}{\bibfnamefont{J.~O'M.}~\bibnamefont{Bockris}},
\bibinfo{author}{\bibfnamefont{A.~K.~N.}~\bibnamefont{Reddy}}, \bibnamefont{and}
  \bibinfo{author}{\bibfnamefont{M.} \bibnamefont{Gamboa-Aldeco}}, 
 \bibinfo{book}{Modern Electrochemistry 2A, Second Edition},
 \bibinfo{publisher}{Kluwer Academic (New York)}
  \bibinfo{pages}{1035-1166} (\bibinfo{year}{2000}).
  
     \bibitem[{\citenamefont{Weast}(2004)}]{Handbook}
 \bibinfo{book}{Handbook of Chemistry and Physics, 58th Edition},
 \bibinfo{publisher}{CRC Press, Inc. (West Palm Beach)} (\bibinfo{year}{1977}).
 
   \bibitem[{\citenamefont{Boyer et~al.}(1994)}]{Boyer_1994}  When the separation $z \ll A$, $C$ is comparable to that of a sphere of radius $A/2$ suspended a distance $z$ above an infinite conducting plane, a problem which results in a logarithmic dependence of $C$ on $z$.  See: 
\bibinfo{author}{\bibfnamefont{L.}~\bibnamefont{Boyer}},
\bibinfo{author}{\bibfnamefont{F.}~\bibnamefont{Houz\'e}},
\bibinfo{author}{\bibfnamefont{A.}~\bibnamefont{Tonck}},
\bibinfo{author}{\bibfnamefont{J.-L.}~\bibnamefont{Loubet}},
\bibnamefont{and}
\bibinfo{author}{\bibfnamefont{J.-M.}~\bibnamefont{Georges}}, 
 \bibinfo{journal}{J. Phys. D: Appl. Phys.} \textbf{\bibinfo{volume}{27}},
  \bibinfo{pages}{1504} (\bibinfo{year}{1994}). 
 
   \bibitem[{\citenamefont{Case}(2007)}]{TBP}
\bibinfo{author}{\bibfnamefont{S.~C.}~\bibnamefont{Case}} (to be published.)
 
  \bibitem[{\citenamefont{Oron et~al.}(1997)}]{Oron_1997}
\bibinfo{author}{\bibfnamefont{A.}~\bibnamefont{Oron}},
\bibinfo{author}{\bibfnamefont{S.~H.}~\bibnamefont{Davis}}, \bibnamefont{and}
  \bibinfo{author}{\bibfnamefont{S.~G.} \bibnamefont{Bankoff}}, 
 \bibinfo{journal}{Rev. Mod. Phys.} \textbf{\bibinfo{volume}{69}},
  \bibinfo{pages}{931} (\bibinfo{year}{1997}).

\end{thebibliography}
\end{document}